\documentclass[11pt,a4paper]{elsart}

\usepackage{ifthen,graphics}
\usepackage{color}
\usepackage{cite}
\usepackage[colorlinks]{hyperref}
\usepackage{epsfig}

\definecolor{dgreen}{rgb}{.0,.7,.0}

\newcommand{\pstar}{\ensuremath{p^*}}
\newcommand{\lc}{\ensuremath{l_\mathrm{c}}}
\newcommand{\dc}{\ensuremath{d_\mathrm{c}}}
\newcommand{\rxy}{\ensuremath{r_{xy}}}
\newcommand{\Emiss}{\ensuremath{E_\mathrm{miss}}}

\newcommand{\dmp}{\ensuremath{\Delta_{\mu \pi}}}
\newcommand{\dep}{\ensuremath{\Delta_{e \pi}}}

\newcommand{\LK}{\ensuremath{L_{K}}}

\renewcommand{\vec}[1]{\mbox{\boldmath{$\rm#1$}}}

\renewcommand{\thesection}{\arabic{section}}%
\makeatletter
\renewcommand{\section}{\@startsection{section}%
{1}%
{0mm}%
{0.5\baselineskip}
{0.3\baselineskip}
{\normalfont\large\bf\mathversion{bold}}}%
\makeatother
\makeatletter
\renewcommand{\subsection}{\@startsection{subsection}%
{2}%
{0mm}%
{0.3\baselineskip}
{0.1\baselineskip}
{\normalfont\normalsize\bf\mathversion{bold}}}%
\makeatother
\makeatletter
\def\ps@copyright{\let\@mkboth\@gobbletwo
  \def\@oddhead{}%
  \let\@evenhead\@oddhead
  \def\@oddfoot{\small\slshape
    \def\@tempa{0}
    \ifx\@volume\@tempa
Preprint submitted to \@journal\hfil\@date\/%
    \else
      Article published in \@jou@vol@pag\hfil\hbox{}\fi}%
  \let\@evenfoot\@oddfoot
}
\makeatother

\def\ifm#1{\relax\ifmmode#1\else$#1$\fi}  
\def\DAF{DA\char8NE}
\def\x{\ifm{\times}}  
\def\pt#1,#2,{\ifm{#1\x10^{#2}}}
\def\up#1{\ifm{^{#1}}}  
\def\dn#1{\ifm{_{#1}}}
\def\ab{\ifm{\sim}}  
  
\def\gam{\ifm{\gamma}} 
\def\to{\ifm{\rightarrow}} 
\def\kl{\ifm{K_L}}   
\def\ks{\ifm{K_S}}
\def\eiii{\ifm{\pi^\pm e^\mp\nu}}  
\def\keiii{\ifm{K_{e3}}}
\def\muiii{\ifm{\pi^\pm \mu^\mp\nu}}   
\def\kmuiii{\ifm{K_{\mu3}}}
\def\pio{\ifm{\pi^0\pi^0}} 
\def\po{\ifm{\pi^0}}
\def\pic{\ifm{\pi^+\pi^-}}  

\def\rmk{\rm\kern.5mm }   
\def\f{\ifm{\phi}}  
\def\Vus{\ifm{|V_{us}|}}   \def\vst{\vphantom{\vrule height4mm depth0pt}}
\def\figbox#1;#2;{\parbox{#2cm}{\epsfig{file=#1.eps,width=#2cm}}}

\newcommand{\aff}[2]
  {Dipartimento di Fisica dell'Universit\`a #1 e Sezione INFN, #2, Italy.}
\newcommand{\affd}[1]
  {Dipartimento di Fisica dell'Universit\`a e Sezione INFN, #1, Italy.}
\begin{document}

\begin{frontmatter}
\title{\mathversion{bold}Measurements of the Absolute Branching Ratios for the 
Dominant \kl\ Decays, the \kl\ Lifetime, and $V_{us}$ with the KLOE Detector}
%
\collab{The KLOE Collaboration}
\author[Na]{F.~Ambrosino},
\author[Frascati]{A.~Antonelli},
\author[Frascati]{M.~Antonelli},
\author[Roma3]{C.~Bacci},
\author[Frascati]{P.~Beltrame},
\author[Frascati]{G.~Bencivenni},
\author[Frascati]{S.~Bertolucci},
\author[Roma1]{C.~Bini},
\author[Frascati]{C.~Bloise},
\author[Roma1]{V.~Bocci},
\author[Frascati]{F.~Bossi},
\author[Frascati,Virginia]{D.~Bowring},
\author[Roma3]{P.~Branchini},
\author[Roma1]{R.~Caloi},
\author[Frascati]{P.~Campana},
\author[Frascati]{G.~Capon},
\author[Na]{T.~Capussela},
\author[Roma3]{F.~Ceradini},
\author[Frascati]{S.~Chi},
\author[Na]{G.~Chiefari},
\author[Frascati]{P.~Ciambrone},
\author[Virginia]{S.~Conetti},
\author[Frascati]{E.~De~Lucia},
\author[Roma1]{A.~De~Santis},
\author[Frascati]{P.~De~Simone},
\author[Roma1]{G.~De~Zorzi},
\author[Frascati]{S.~Dell'Agnello},
\author[Karlsruhe]{A.~Denig},
\author[Roma1]{A.~Di~Domenico},
\author[Na]{C.~Di~Donato},
\author[Pisa]{S.~Di~Falco},
\author[Roma3]{B.~Di~Micco},
\author[Na]{A.~Doria},
\author[Frascati]{M.~Dreucci},
\author[Frascati]{G.~Felici},
\author[Karlsruhe]{A.~Ferrari},
\author[Frascati]{M.~L.~Ferrer},
\author[Frascati]{G.~Finocchiaro},
\author[Frascati]{C.~Forti},
\author[Roma1]{P.~Franzini},
\author[Frascati]{C.~Gatti},
\author[Roma1]{P.~Gauzzi},
\author[Frascati]{S.~Giovannella},
\author[Lecce]{E.~Gorini},
\author[Roma3]{E.~Graziani},
\author[Pisa]{M.~Incagli},
\author[Karlsruhe]{W.~Kluge},
\author[Moscow]{V.~Kulikov},
\author[Roma1]{F.~Lacava},
\author[Frascati]{G.~Lanfranchi},
\author[Frascati,StonyBrook]{J.~Lee-Franzini},
\author[Karlsruhe]{D.~Leone},
\author[Frascati]{M.~Martini},
\author[Na]{P.~Massarotti},
\author[Frascati]{W.~Mei},
\author[Na]{S.~Meola},
\author[Frascati]{S.~Miscetti},
\author[Frascati]{M.~Moulson},
\author[Karlsruhe]{S.~M\"uller},
\author[Frascati]{F.~Murtas},
\author[Na]{M.~Napolitano},
\author[Roma3]{F.~Nguyen},
\author[Frascati]{M.~Palutan},
\author[Roma1]{E.~Pasqualucci},
\author[Roma3]{A.~Passeri},
\author[Frascati,Energ]{V.~Patera},
\author[Na]{F.~Perfetto},
\author[Roma1]{L.~Pontecorvo},
\author[Lecce]{M.~Primavera},
\author[Frascati]{P.~Santangelo},
\author[Roma2]{E.~Santovetti},
\author[Na]{G.~Saracino},
\author[Frascati]{B.~Sciascia},
\author[Frascati,Energ]{A.~Sciubba},
\author[Pisa]{F.~Scuri},
\author[Frascati]{I.~Sfiligoi},
\author[Frascati,Novo]{A.~Sibidanov},
\author[Frascati]{T.~Spadaro},
\author[Roma1]{M.~Testa},
\author[Roma3]{L.~Tortora},
\author[Roma1]{P.~Valente},
\author[Karlsruhe]{B.~Valeriani},
\author[Frascati]{G.~Venanzoni},
\author[Roma1]{S.~Veneziano},
\author[Lecce]{A.~Ventura},
\author[Roma3]{R.Versaci},
\author[Frascati,Beijing]{G.~Xu}

\address[Beijing]{Permanent address: Institute of High Energy 
Physics of Academica Sinica,  Beijing, China.}
\address[Frascati]{Laboratori Nazionali di Frascati dell'INFN, 
Frascati, Italy.}
\address[Karlsruhe]{Institut f\"ur Experimentelle Kernphysik, 
Universit\"at Karlsruhe, Germany.}
\address[Lecce]{\affd{Lecce}}
\address[Moscow]{Permanent address: Institute for Theoretical 
and Experimental Physics, Moscow, Russia.}
\address[Na]{Dipartimento di Scienze Fisiche dell'Universit\`a 
``Federico II'' e Sezione INFN,
Napoli, Italy}
\address[Novo]{Permanent address: Budker Institute of Nuclear Physics, Novosibirsk, Russia.}
\address[Pisa]{\affd{Pisa}}
\address[Energ]{Dipartimento di Energetica dell'Universit\`a 
``La Sapienza'', Roma, Italy.}
\address[Roma1]{\aff{``La Sapienza''}{Roma}}
\address[Roma2]{\aff{``Tor Vergata''}{Roma}}
\address[Roma3]{\aff{``Roma Tre''}{Roma}}
\address[StonyBrook]{Physics Department, State University of New 
York at Stony Brook, USA.}
\address[Virginia]{Physics Department, University of Virginia, USA.}

\corauth[cor1]{cor1}{\small $^1$ Corresponding author: Mario Antonelli
INFN - LNF, Casella postale 13, 00044    Frascati (Roma), 
Italy; tel. +39-06-94032728, e-mail Mario.Antonelli@lnf.infn.it}

\begin{abstract}\vglue-3mm
\noindent  From a sample of about $10^9$ \f\ mesons produced at \DAF, we have selected
  \kl\ mesons tagged by observing \ks\ \to\ \pic\ decays.
  We present results on the major \kl\ branching ratios, including
  those of the semileptonic decays needed for the determination of \Vus.
  These branching ratio measurements are fully inclusive with respect 
  to final-state radiation.
  The \kl\ lifetime has also been measured.
\end{abstract}
\end{frontmatter}
\section{Introduction} 
We have measured the absolute branching ratios (BR's) 
for \kl\ decays to \eiii,\/ \muiii,\/ \pic\po,\/ and \pio\po,\/ as well as the 
lifetime of the \kl\ meson.
The branching ratios and lifetime are necessary for the determination of
the partial rates for semileptonic \kl\ decays, from which the CKM matrix 
element \Vus\ can be obtained.
Absolute branching ratios are not usually measured. 
The PDG derives the \kl\ BR's from a ``fit'' to ratios of partial 
widths \cite{PDG}, giving errors in the 0.5--2\% range.
Recently, KTeV \cite{KTeV:BrL}, measuring a complete set of such ratios 
and requiring that all BR's sum to unity, has obtained new BR values in 
strong disagreement with the PDG values.
The NA48 value for BR(\kl\dn{,\,e3}) \cite{NA48:ke3} essentially agrees with 
that from KTeV, though the NA48 measurement is sensitive to
${\rm BR}(\pio\po)$ and in fact makes use of the KTeV value for the 
latter, averaged with older measurements.

 Measurements of the absolute \kl\ branching ratios are a unique
 possibility of the \f\ factory, where kaons are produced in $\f\to\ks\kl$ 
 decays. A pure sample of nearly
 monochromatic\footnote{\f\ mesons at \DAF\ are produced with $p_{\f}\ab12$ 
 MeV/$c$, toward the ring's center.} \kl's
 can be selected by identification of \ks\ decays (tagging).  The absolute
 branching ratios can be determined by counting the fraction of \kl's 
 that decay into each channel and
 correcting for acceptances, reconstruction efficiencies, and background.
 In addition, the count $N$ of all decays in a time interval 
$\Delta t=t_2-t_1$ gives the lifetime with a fractional 
error equal to 1/$\sqrt N$, independently of the value of $\Delta t$. 
In this work, we use a slightly different approach. 
We begin by assuming the \kl\ lifetime to be known and equal to the PDG value,
and measure the number of decays in each of the major channels to obtain
the corresponding BR's. We then observe that the sum of the BR's 
differs slightly from one. Renormalizing
the sum to one, we find the correct lifetime and BR values.

\section{Tagging uncertainties}
\label{sec:tag1}
The efficiency for identification of the tagging $\ks\to\pic$ decay
depends slightly on the fate of the \kl: it is different for events 
in which the \kl\ decays to each channel, interacts in the calorimeter,
or escapes the detector.
We define the {\em tag bias} for the detection of \kl\ decays to a state 
$f$ as the ratio of the tagging efficiency, $\epsilon_{{\rm tag},\,f}$,
  for events in which \kl\ \to $f$ within the fiducial volume (FV) to 
 the overall tagging efficiency, $\bar{\epsilon}_{\rm tag}$,
 determined without regard to the fate of the \kl. 
 The tag bias for channel $f$ is thus 
 $\epsilon_{{\rm tag},\,f}\,/\,\bar{\epsilon}_{\rm tag}$, and
\begin{equation}
{\rm BR}(\kl\to f) = 
\frac{N_f}
     {N_{{\rm tag}, f}\,\epsilon_{\rm FV}\,\epsilon_{{\rm rec}, f}}\cdot
\frac{\bar{\epsilon}_{\rm tag}}{\epsilon_{{\rm tag},f}},
  \label{eqn:brkl_master}
\end{equation}
where $N_f$ is the number of \kl\ decays to $f$, 
$N_{\rm tag}$ is the number of tagged \kl's,  $\epsilon_{\rm FV}$ is
the fraction of decays in the FV, and 
$\epsilon_{{\rm rec},\,f}$ is the reconstruction efficiency.
Losses of \kl's from interactions in the beam pipe, chamber walls, and
chamber gas are 
taken into account in the evaluation of $\epsilon_{\rm FV}$.

We use a FV within the chamber defined by $35<\rxy<150$ cm and 
$|z|<120$ cm, where $(x,y,z)$ are the coordinates of the \kl\ decay 
vertex.\footnote{The $z$-axis coincides with the bisectrix of the 
two beams.} 
The mean \kl\ decay length in KLOE is 340 cm and \ab26.1\% of the \kl's decay 
in the FV. The dependence of the FV efficiency on the \kl\ lifetime is
\begin{equation}
\epsilon_{\rm FV}/\epsilon_{\rm FV}^0 = 
1 + 0.0128\ {\rm ns}^{-1}\,(\tau^0 - \tau)
\label{eq:FVtau}
\end{equation}
for values of $\tau_{\kl}$ around 50 ns (here, $\epsilon_{\rm FV}^0$ refers to
the FV efficiency calculated using $\tau^0 = 51.7$~ns).

\section{The KLOE detector}
\label{sec:kloedet}
The KLOE detector consists of a large cylindrical 
drift chamber (DC), surrounded by a 
lead/scintillating-fiber electromagnetic calorimeter 
(EMC). A superconducting coil around the calorimeter 
provides a 0.52 T field. 
The drift chamber \cite{KLOE:DC} is
4~m in diameter and 3.3~m long.
The momentum 
resolution is $\sigma_{p_{\perp}}/p_{\perp}\approx 0.4\%$. 
Two-track vertices are reconstructed with a spatial 
resolution of \ab\ 3~mm. 
The calorimeter \cite{KLOE:EmC} 
is divided into a barrel and two endcaps. It covers 98\% of the solid 
angle. Cells close in time and space are grouped into 
calorimeter clusters. The energy and time resolutions 
are $\sigma_E/E = 5.7\%/\sqrt{E\ {\rm(GeV)}}$ and  
$\sigma_T = 54\ {\rm ps}/\sqrt{E\ {\rm(GeV)}}\oplus50\ {\rm ps}$, 
respectively.
The KLOE trigger \cite{KLOE:trig} 
uses calorimeter and chamber information. For this 
analysis, only the calorimeter signals are used. 
Two energy deposits above threshold ($E>50$ MeV for the 
barrel and  $E>150$ MeV for the endcaps) are required. 
Recognition and rejection of cosmic-ray events is also 
performed at the trigger level. Events with two energy 
deposits above a 30 MeV threshold in the outermost 
calorimeter plane are rejected.

\section{Data and Monte Carlo samples}
\label{sec:datasample}

For the present measurements, we use a subset of the data collected by 
KLOE at \DAF\ during the years 2001 and 2002 that satisfies basic quality 
criteria \cite{KLOEnote}. The data used corresponds to an integrated 
luminosity of \ab328 pb$^{-1}$.

Each run used in the analysis is simulated with the KLOE Monte Carlo (MC) 
program, \textsc{GEANFI} \cite{KLOE:offline}, using values of relevant machine 
parameters such as $\surd s$ and $\vec p_\f$ as determined from data. 
Machine background obtained from the data is superimposed on MC events
on a run-by-run basis. The number of events simulated for each run in the 
data set is equivalent to that expected on the basis of the run luminosity. 
For this analysis, we use an MC sample consisting of $\f\to\ks\kl$ events
in which the \ks\ and \kl\ decay in accordance with their natural 
branching ratios. The effects
of initial- and final-state radiation are included in the simulation. 
The treatment of final-state radiation in \ks\ and \kl\ decays 
is discussed in Ref.~\citen{KLOE:gatti}. In particular, the fraction 
of \keiii\ events produced by the generator for various cutoff values
for the energy and angle of the radiated photon compares well 
with the results of Ref.~\citen{andre}.

Good runs, and the corresponding MC data, are 
organized into 14 periods.
The branching ratio analysis is performed independently for each of these
periods. Averages over the 14 periods give the final results.
One-fourth of the data is used to compute various
corrections; the remainder is used for for the evaluation of the 
\kl\ branching ratios.
This choice minimizes the total uncertainty. 

\section{\kl\ tag by $\ks\to\pic$}\label{sec:event}

$\ks\to\pic$ decays are identified by two tracks of opposite curvature from a vertex
in a cylindrical FV with $\rxy < 10$~cm and
$|z| < 20$~cm, 
centered on the collision region as determined for each run using
Bhabha events.
We also require that the two tracks give $|m(\pic)-m_{K^0}|<5$ MeV/$c^2$
and that 
$|\sum(\vec p_++\vec p_-)|-p_K<$ 10 MeV/$c$, with $p_K$ calculated from
the kinematics of the $\f\to\ks\kl$ decay.
If more than one vertex is found within the FV, that closest to the nominal 
position of the IP is used.
The \kl\ momentum, $p_{\kl}$, 
is obtained from the \ks\ direction and $\vec p_\f$. 
The resolution is \ab0.8 MeV/$c$ and is dominated by the beam-energy spread.
The position of the \f\ production point, $\vec x_\f$, is taken as
the point of closest approach of the \ks\ direction to the beam line.
The \kl\ line of flight is then constructed from the \kl\ momentum, 
$\vec p_{\kl} = \vec p_\f-\vec p_{\ks}$, and the position of the production 
vertex, $\vec x_\f$.

The tag bias is mostly due to the dependence of the calorimeter trigger 
efficiency on the behavior of the \kl\ and is estimated by MC to vary from 
0.99 to 1.06 for \kl\ decays to charged particles and to $\pio\po$. 
However, poor knowledge of the cross section for \kl\ interactions in the 
calorimeter limits the accuracy of the simulation. 
To decrease tag-bias effects, we require that the \ks\ by itself satisfy 
the calorimeter trigger. This is accomplished by requiring that there 
be two clusters associated to the tracks from the $\ks\to\pic$ decay.
The clusters must be in fired trigger sectors in the calorimeter barrel, 
and must have energies above a 100 MeV threshold.
Tag bias is also introduced by the dependence of the $\ks\to\pic$ 
reconstruction efficiency on the presence of other tracks in the chamber 
(``track reconstruction interference'').
Symmetric \ks\ decays produce less interference and so we require 
$|p_+-p_-|<60$~MeV/$c$ for \ks\ decays.
After imposing these requirements, we are left with a tagging efficiency 
of about 9\%. 
The tag bias for \kl\ decays to charged particles is about 0.97 and for 
decays to \pio\po\ about 1.00.  

The MC estimate of the tagging efficiency is corrected using data.
$\bar{\epsilon}_{\rm tag}$ has been corrected by about $+0.2$\% to account 
for the effect of the cosmic-ray veto in the trigger, where this correction is 
obtained from a downscaled sample of events for which the veto was
not enforced. An additional correction of about $-0.5$\% accounts for a small 
data-MC disagreement in the effect of track reconstruction interference.
These corrections are run-period dependent.

\section{\kl\ decays into charged particles}
\label{sec:charged}

All tracks in the chamber, after removal of those from 
the \ks\ decay and their descendants, are extrapolated to their
points of closest approach to the line of flight of the \kl, $\vec x_{\rm c}$. 
The distance of closest approach, \dc, 
the momentum $\vec p_{\rm c}$ of the track at $\vec x_{\rm c}$,
and the extrapolation length \lc\ are then computed. 
For each sign of charge, we consider
the track with the smallest value of \dc\ to be associated to the \kl\ decay.
Candidate tracks from \kl\ decays must satisfy
$\dc< a \rxy + b$, with $a=0.03$ and $b=3$ cm,
and $-20$ cm $<\lc<25$ cm.

For the selection of \kl\ decays into charged particles, we require two
good (i.e., satisfying the above criteria) \kl\ decay tracks of opposite 
sign of charge.
To improve the reconstruction, we then apply a vertex-reconstruction 
algorithm.
The average efficiency for complete reconstruction is 
54\% for \keiii, 52\% for \kmuiii, and 38\% for \pic\po,
as evaluated from MC simulation. 
These efficiencies are then corrected as described in the following;
the corrections range between 0.99
and 1.03 depending on the channel and on the run period.

 The vertex efficiency is about 97\%,
 and the values obtained from the MC are in agreement with those
 for data within 0.1\%.
 The tracking efficiency is determined by counting the number of
 events in which there is a \kl\ decay track of at least one sign
 and the number of events in which there are two decay tracks of 
 opposite sign. The data-MC difference in the single-to-double ratio 
 is then used to correct the MC estimate for the tracking efficiency.
 This correction is evaluated as function of the track momentum using 
 $\kl\to\pic\po$ and \eiii\ events. 
 
 To select $\kl\to\pic\po$ events, we require at least 
 one good \kl\ decay track and two photons from a \po\ decay.
 To reject background, cuts are applied on the two-photon invariant mass
 and on the times of flight of the photons \cite{KLOEnote}.
 The momentum of one
 decay track is calculated with a resolution of about 10 MeV/$c$
 using the two photons from the \po\ decay and the momentum of the
 other decay track.
 In order to obtain the correction for higher momentum tracks
 (above $\sim$ 150 MeV/$c$), \eiii\ decays are used.  
 $\kl\to\eiii$ decays are selected 
 by identifying the electrons by time of flight with a purity of \ab95\%.
 The remaining background is due mostly to $K_{\mu 3}$ decays.
 The momentum of the second \kl\ track is evaluated from the missing 
 energy with a resolution of about 30 MeV/$c$. 
 The comparison of the tracking efficiency obtained from
 $\kl\to\eiii$ events and that obtained from 
 $\kl\to\pic\po$ events shows
 good agreement in the overlap region, $100 < p < 150$~MeV/$c$,
 as seen from Fig.~\ref{fig:efftrk_over}.
\begin{figure}[hbt]
  \centerline{\figbox efftrk_overlap_dt;6.5;\kern1cm
              \figbox efftrk_overlap_ratio;6.5 ;}
  \caption{Left: Tracking efficiency from 
    $\kl\to\pic\po$ (dotted line) 
    and $\kl\to\eiii$ (full line) events, 
    for data. Right: Data-MC ratio of tracking efficiencies 
    for the two samples.}
  \label{fig:efftrk_over}
\end{figure}
 Because the expected momentum is determined with better precision
 in $\kl\to\pic\po$ events at low momentum, 
 for momenta smaller than 150 MeV/$c$, we use the correction obtained 
 from this sample, while for momenta greater than 150 MeV/$c$, we use the 
 correction obtained from $\kl\to\eiii$ events.

 MC studies have shown that the best discriminant 
 amongst the \kl\ decay modes into charged particles is the smaller
 absolute value of the two possible values of 
$\dmp = |\vec p_{\rm miss}| - \Emiss$, where 
$\vec p_{\rm miss}$ and \Emiss\ are 
 the missing momentum and energy at the \kl\ vertex, evaluated
 by assigning one track the pion mass and one track the muon mass.
 The shape of the \dmp\ spectrum is very sensitive to the radiative
 corrections to \kl\ decay processes. This 
 is particularly evident for \keiii\ decays (see Fig.~\ref{fig:kell3}).

\begin{figure}[hbt]
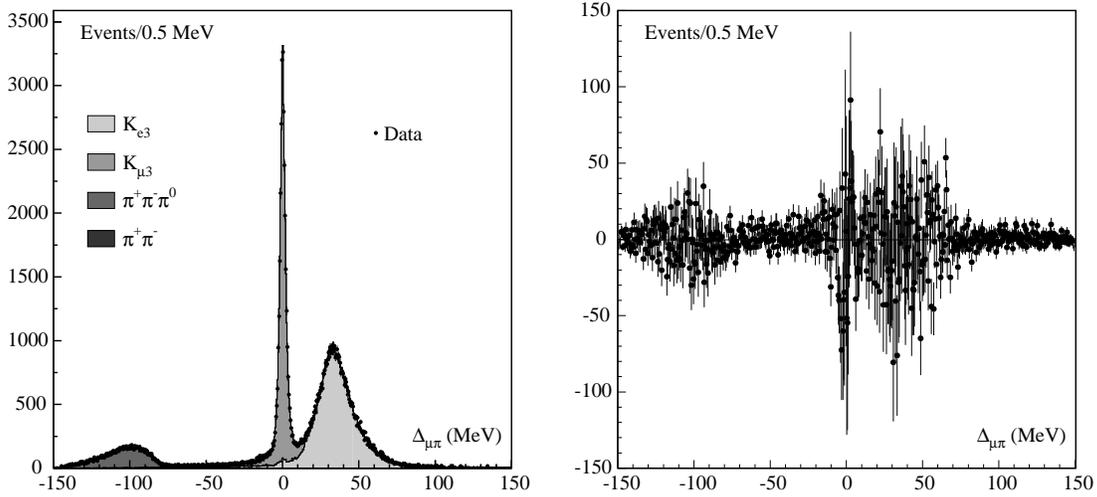

  \centerline{\figbox spectra;7.;\kern5mm\figbox spedif;7.;}
  \caption{Distribution of \dmp\ for a single run period, 
    with fit to MC distributions for different decay channels.}
  \label{fig:fit_example}
\end{figure}

The numbers of $\kl\to\eiii$, $\kl\to\muiii$, and $\kl\to\pic\po$
events, fully inclusive with respect to final-state radiation, 
are obtained from fits to the \dmp\
distributions with 
a linear combination of MC distributions.
An example of such a fit is shown in Fig.~\ref{fig:fit_example}.

 The residual background consists of
 of $\kl\to\pic$ decays, $\kl\to\pio\po$ decays with $\po \to e^+ e^- \gamma$,
 $\kl\to\ks\to\pic$ regeneration, and $\Lambda/\,\Sigma$ production
 from \kl-nuclear interactions. The total amount of background is about 0.5\%.

In the MC, the momentum resolution is slightly better than it is for data.
As described in Ref.~\cite{KLOEnote}, reconstructed track momenta for MC
events are slightly smeared to obtain better agreement with data.

\section{$\kl\to\pio\po$ decays}
\label{sec:neutral}

The position of the \kl\ vertex for $\kl\to\pio\po$ decays is obtained from
the arrival times of the photons at the calorimeter.
The method is fully described in Refs.~\citen{KLOE:EmC}
and~\citen{KLOE:offline}.
The position of the \kl\ vertex is assumed to be along 
the \kl\ line of flight as reconstructed from the tagging $\ks\to\pic$ decay. 
The arrival time of each photon gives an independent determination of
the \kl\ path length, \LK, up to a two-fold ambiguity. 
One solution is easily discarded. 
The final value of \LK\ is obtained from the weighted average 
of the different measurements.
The accuracy of the method is checked using $\kl\to\pic\po$ decays,
by comparing the position of the \kl\ decay vertex from the reconstruction 
of the tracks from the \pic\ pair with the position obtained from photon 
arrrival-time measurements.
Decays to \pic\po\ are also used to measure the single-photon efficiency,
for both data and MC.

For the selection of $\kl\to\pio\po$ events, we require at least 
three photons with $E>20$~MeV originating from the \kl\ decay point.
The background contamination, 
which is concentrated in the three- and four-photon event samples,
is dominated by $\kl\to\pic\po$ events with additional clusters 
due to machine background. 

Additional cuts are therefore made for events with three and four photons.
First, we require that at least one of the 
photons have $E > 50$ MeV. We then select 
the pair of clusters with the smallest separation,
and reject events which satisfy
$
  E_{\rm min} < [70 + 250(|\cos\theta | - 1)]~\mbox{MeV}
$
and
$
  |\cos\theta| > 0.9
$,
where $E_{\rm min}$ is the energy of the lower-energy cluster 
and $\theta$ is the polar angle of the mean position of the pair of clusters.
These cuts are effective at reducing the contamination from 
$\kl\to\pic\po$ events in which an extra cluster is present due to 
machine background.
Moreover, the RMS pull of the individual $L_K$ measurements about the
weighted-average value is required to be smaller than 1.2 in events with 
three photons.
Finally, tracks from decays of the \kl\ into charged 
secondaries are found to have comparatively 
high values of \pstar, the magnitude of the track momentum in the rest 
frame of the \kl. We therefore reject three-photon events in which 
$
 \pstar c > E_{\rm tot} - 114~\mbox{MeV}
$, where $E_{\rm tot}$ is the sum of the photon energies.
This cut eliminates background from \kl\ decays into charged particles
without incurring inefficiencies in the presence of tracks from
machine background, photon conversions, or unassociated track segments 
from the \ks\ decay.

After making these selection cuts, we obtain the results listed in 
Table \ref{tab:select_resu} for the photon multiplicity distribution for 
vertices in the fiducial volume.
The distributions for data and MC agree at 
the 0.5\% level.
\begin{table}[hbt]
  \begin{center}
\renewcommand{\arraystretch}{1.1} 
    \begin{tabular}{ccc}
      \hline\hline
\vst$n$ & $N_{n\gamma}/N_{\rm tot}$, data & $\Delta(N_{n\gamma}/N_{\rm tot})$ data$-$MC\\
      \hline
\vst  3 & $(2.22 \pm 0.02)\%$  & $(0.41 \pm 0.02)\times 10^{-2}$ \\
      4 & $(9.05 \pm 0.05)\%$  & $(0.18 \pm 0.06)\times 10^{-2}$ \\
      5 & $(32.09 \pm 0.05)\%$ & $(0.49 \pm 0.08)\times 10^{-2}$ \\
      6 & $(54.71 \pm 0.08)\%$ & $(-0.51 \pm 0.1)\times 10^{-2}$ \\
      7 & $(1.82 \pm 0.02)\%$  & $(0.36 \pm 0.02)\times 10^{-2}$ \\
      \hline
    \end{tabular}
  \end{center}
  \caption{Comparison of photon multiplicity distributions for data and MC 
  after all selection cuts and corrections. The values for 
  the MC refer to the combined signal and background samples.}
  \label{tab:select_resu}
\end{table}
The reconstruction efficiency $\epsilon_{\rm rec}$ and the purity 
have been evaluated for each of the 14 data-taking periods.
Their average values are $\sim 99.0\%$ and $98.9\%$, respectively. 

\section{Systematic uncertainties}
\label{sec:sys}
The systematic uncertainties on the absolute BR measurements 
 are summarized in Table~\ref{tab:sys} for each \kl\ decay mode.
Correlations among channels are taken into account, and the
complete information is contained in the error matrix \cite{KLOEnote}.
\begin{table}[hbt]
\renewcommand{\arraystretch}{1.1} 
  \begin{center}
  \begin{tabular}{lcccc}\hline\hline
\vst& \eiii & \muiii   & \pio\po & \pic\po \\\hline
\vst
              Tag bias  & 0.0036 & 0.0034   & 0.0029 & 0.0058 \\
$\epsilon_{\rm FV}$     & 0.0058 & 0.0058   & 0.0058 & 0.0058 \\
 Selection efficiencies & 0.0028 & 0.0026   & 0.0100 & 0.0032 \\
       Kinematic shape  & 0.0015 & 0.0033   & -      & 0.0079 \\\hline
  \end{tabular}
  \end{center}
  \caption{Summary of the fractional 
  systematic uncertainties on the absolute BR
 measurements.}
\label{tab:sys}
\end{table}

The systematic error on the evaluation of the tag bias 
is taken from the uncertainties on the individual corrections 
and from the study of the stability of the results obtained with 
different tagging criteria. The values of the cuts used to define the tag
have been varied over a wide range.
In total, we have studied the stablity of the result for 15
different tag configurations.
In this study, the tagging efficiency 
changes by $\sim\pm$50\% with respect to the tagging efficiency
obtained with the default tag.
Corresponding fractional changes on the \kl\ branching ratios of 
0.25\% to 0.56\% are observed depending on the channel 
(see Table~\ref{tab:sys}, line 1). 
We have also determined the tag bias with the requirement 
that the pions from $\ks\to\pic$ be associated to calorimeter clusters in
unoccupied trigger sectors.
The fractional change in the result with respect to 
that obtained in the default configuration is smaller than 0.05\%.
Finally, a further 0.15\% systematic uncertainty that is fully 
correlated between decay modes
 arises
 from the uncertainty on the correction for
 the track reconstruction interference effect.
The effects of the limited knowledge of the \kl\ lifetime
and the nuclear interaction cross section
 are found to be negligible.

The main sources of uncertainty on the evaluation of the fiducial-volume
efficiency are the limited knowledge of the \kl-nuclear interaction cross 
section and the accuracy of the \kl\ lifetime value. 
The uncertainty on the fiducial-volume efficiency
obviously affects the BR measurements for all channels in the 
same way and its contribution cancels out 
if the sum of the branching ratios is constrained to unity. 
 The 0.85\% current accuracy 
on the \kl\ lifetime \cite{PDG} leads to a 0.56\% uncertainty on the 
absolute BR measurements (Table~\ref{tab:sys}, line 2).
 The \kl\ beam loss due to nuclear interactions in the beam pipe and 
inner DC wall is estimated with the MC and corrected 
using data, as described in Ref.~\citen{KLOEnote}.
The corrections are determined with an uncertainty of about 30\%,
both for regeneration and for $\Lambda/\Sigma$ production, resulting
in a 0.15\% relative contribution to the uncertainty on the fiducial-volume 
efficiency.

The uncertainty on the tracking efficiency correction is 
determined from sample statistics and
by the variation of the results observed using different criteria
to identify good tracks from \kl\ decays.
The fractional statistical uncertainty amounts to about 0.1\% for 
\keiii\ and \kmuiii\ decays, and to about 0.3\% for \pic\po\ decays. 
The effect of differences in the resolution with which the variables
\dc\ and \lc\ (Sec.~\ref{sec:charged}) are reconstructed
in data and in MC events and the possible bias introduced
 in the selection of the control sample have been studied by varying 
the values of the cuts made when associating tracks to \kl\ vertices.
These changes result in a variation of the tracking efficiency
of about $\pm$20\%.
Corresponding fractional changes of the \kl\ branching ratios of 
0.26\% to 0.32\% are observed, depending on the channel 
(Table~\ref{tab:sys}, line 3).

\begin{figure}[hbt]
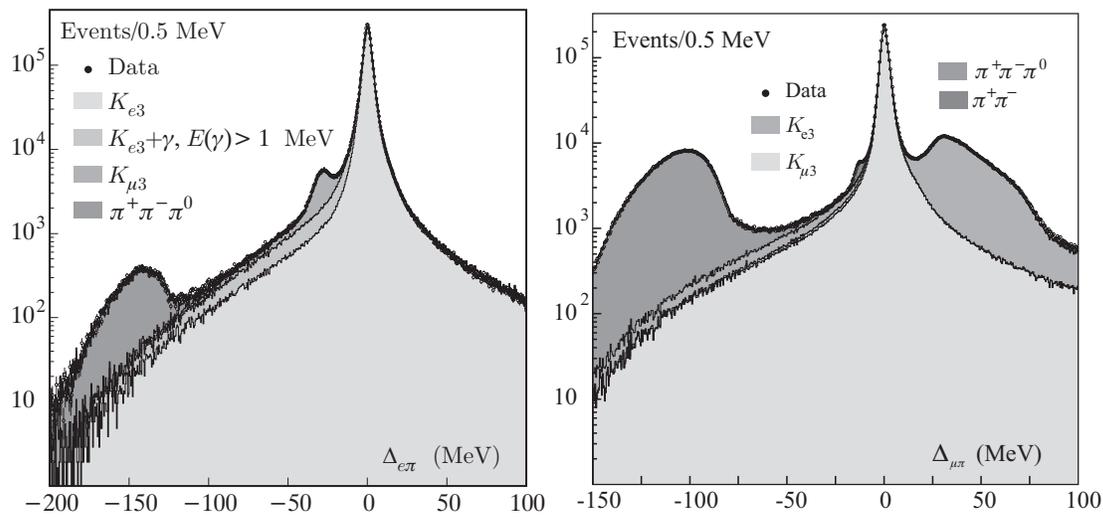

  \centerline{\figbox fig1-c;7.1;\kern1mm\figbox fig2-c;7.3;}
  \caption{Left: Distribution of \dep\ for a sample enriched in \keiii\
    events using electron identification by time-of-flight and energy
    deposition measurements, for data and MC. The contribution from 
    $\keiii\gam$ events with $E_\gam>1$~MeV is also shown. Right: 
    Distribution of \dmp\ for a \kmuiii-enriched sample.}
\label{fig:kell3}
\end{figure}

To assess the uncertainty
on the BR measurements
arising from limited knowledge of the momentum 
resolution and of the accuracy of the simulation of final-state radiation,
we have examined the
agreement between the $|\vec p_{\rm miss}| - \Emiss$ 
 distributions for data and MC
 using \keiii- and \kmuiii-enriched samples selected via
electron identification with the calorimeter \cite{KLOEnote}.
The comparisons are shown in Fig.~\ref{fig:kell3} (\keiii, left,
and \kmuiii, right).
The uncertainties have been estimated by studying the change 
in the results for the BR's with and without 
shape corrections obtained from the data-MC ratios of the
distributions of $|\vec p_{\rm miss}| - \Emiss$ for the control samples,
and by repeating the measurement using 
different mass hypotheses.
In all, five configurations have been studied: 
no shape corrections, 
shape corrections for all modes, 
shape corrections only for \keiii, 
shape corrections only for \kmuiii,
use of the alternative $\pi e$ mass hypothesis.
 Fractional variations of 0.15\%, 0.3\%, and 0.8\% are observed
 for the \keiii,  \kmuiii, and \pic\po\ BR's
 (Table~\ref{tab:sys}, line 4).

The systematic uncertainty on the selection efficiency for $\kl\to\pio\po$
decays includes the uncertainties in the photon-efficiency 
corrections, the effects of the cluster-splitting recovery algorithm, 
and the level of background from nuclear interactions.
These uncertainties have been evaluated by checking the stability of the
measured value of the BR with respect to modifications in the 
selection criteria.
The largest deviation occurs if the minimum number of photons
required at the \kl\ vertex is rasied from three to five.
This cut reduces the background contamination to about 0.1\% while 
maintaining a signal efficiency of 88\%. 
Requiring $n \geq 5$ changes the result by 1\%.
This variation has been conservatively taken as the systematic error
on the measurement of BR($\kl\to\pio\po$) (Table~\ref{tab:sys}, line 3).
The stability of the result for harder photon-energy thresholds
has also been studied. The BR is stable to within 0.3\%
when the energy threshold for single photons is raised from 20~MeV to
50~MeV.

\section{Results and determination of \Vus}

\subsection{Absolute \kl\ branching ratios}
\label{sec:abbr}
A total of about \pt 13,6, tagged \kl\ events are used for the
measurement of the branching ratios. 
The absolute branching ratios are obtained separately for each 
of the 14 run periods.
Good stability of the results is observed once the efficiency
corrections for each run period are applied. 
We first use the value of the \kl\ lifetime 
$\tau_{\kl} = 51.54 \pm 0.44$~ns \cite{vos72} 
to evaluate the fiducial-volume efficiency.
The values for the absolute branching ratios obtained from the 
averages over all run periods are given in Table~\ref{tab:rawbr}.
\begin{table*}[hbt]
\begin{center}
\renewcommand{\arraystretch}{1.1} 
\begin{tabular}{@{}lccccc}
\hline\hline
\vst Mode &   BR   & $\delta$ stat & $\delta$ syst-stat & $\delta$ syst & $\delta$ corr-syst \\ \hline
\vst\eiii & 0.4049 & 0.0006      & 0.0008         & 0.0018        & 0.0025  \\
\muiii &0.2726 & 0.0006      & 0.0006         & 0.0014        & 0.0017 \\ 
\pio\po &0.2018 & 0.0003     & 0.0004         & 0.0023        & 0.0012 \\
\pic\po &0.1276 &0.0004      &  0.0004        & 0.0014         & 0.0008 \\ \hline
\end{tabular}\\[2pt]
\caption{Branching ratios and errors with no constraint on the sum.}
\label{tab:rawbr}
\end{center}
\end{table*}

For decays into charged particles, the statistical errors above are obtained 
from the fit and include a contribution from MC statistics which is 
about equal to that from the signal statistics.
The systematic-statistical error reflects the statistical uncertainties 
in the determination of the various corrections.
The correlated-systematic error reflects all contributions to the errors
on the BR's that are 100\% correlated for all channels, such as 
the uncertainty in the value of the \kl\ lifetime.
The sum of the four BR's above, plus the sum of the PDG values for
\kl\ decays to $\pic$, $\pio$, and $\gam\gam$ (${\rm sum} = 0.0036$) 
\cite{PDG}, 
is $1.0104\pm0.0018_{\rm stat}\pm0.0074_{\rm syst}$.

\subsection{\kl\ branching ratios and lifetime} 
The correct value of the \kl\ lifetime is the value for which the sum 
above is unity. We find
\begin{equation}
  \tau_{\kl} = 50.72 \pm 0.11_{\rm stat} \pm 0.13_{\rm syst-stat}\pm 
  0.33_{\rm syst}~\mbox{ns},
  \label{eqn:lifetime}
\end{equation}
in agreement with the recent direct measurement obtained by KLOE using a 
high-statistics sample of $\kl\to\pio\po$ decays \cite{KLOE:KLlife}.
By renormalizing the sum of the measured BR's to $1-0.0036=0.9964$,
we finally obtain the results in Table~\ref{tab:finalbr}.
\begin{table*}[hbt]
\begin{center}
\renewcommand{\arraystretch}{1.1} 
\begin{tabular}{@{}lcccc}
\hline\hline
\vst Mode &   BR   & $\delta$ stat & $\delta$ syst-stat & $\delta$ syst  \\ \hline
\vst\eiii & 0.4007 & 0.0005        & 0.0004             & 0.0014         \\
\muiii &0.2698 & 0.0005        & 0.0004             & 0.0014         \\ 
\pio\po &0.1997 & 0.0003       & 0.0004             & 0.0019         \\
\pic\po &0.1263 &0.0004        & 0.0003             & 0.0011         \\ \hline
\end{tabular}\\[2pt]
\caption{Final values for branching ratios and errors, with the constraint 
on the sum described in the text.}
\label{tab:finalbr}
\end{center}
\end{table*}

After normalization, the complete correlation matrix is:
\\[1mm]
\begin{displaymath}
\left(
\begin{array}{ccccc}
 1 & -0.25 & -0.56 & -0.07 & 0.25 \\
   &     1 & -0.43 & -0.20 & 0.33 \\
   &       & 1     & -0.39 & -0.21 \\
   &       &       & 1     & -0.39 \\
   &       &       &       & 1     
\end{array}
\right)
\end{displaymath}
where the columns and rows refer, in order, to the errors on the
BR's for decays to $\eiii$, $\muiii$, $\pio\po$, and
$\pic\po$, and to the error on $\tau_{\kl}$.

The ratio $R_{\mu,\,e}=\Gamma(K_{\mu 3}$)/$\Gamma(K_{e3}$) can be computed 
from the value of the form-factor slope $f_0$. From our measurements above, 
we obtain $R_{\mu,\,e}=0.6734\pm0.0059,$ to be compared with the value
$R_{\mu,e}=0.6640\pm0.0040$ calculated using the average of the
measurements of $f_0$ from KTeV \cite{ktev} and ISTRA+ \cite{istra_mu}.
The ratio ${\rm BR}(\kl\to\pio\po)/\,{\rm BR}(\kl\to\pic\po)$ 
can be estimated from the 
isospin amplitudes obtained from branching ratios and Dalitz-plot slopes 
for $K\to3\pi$ decays. We find
\begin{displaymath} 
R_{3\pi}=\mbox{BR}(\kl\to\pio\po)/\mbox{BR}(\kl\to\pic\po)=1.582\pm 0.027,
\end{displaymath}
to be compared with the result of Ref.~\citen{binenz}, $R_{3\pi}=1.579$.

\subsection{Determination of \Vus}  
Measurements of $|V_{us}|$ and $|V_{ud}|$ provide the most precise test of
the unitarity of the CKM mixing matrix, as $|V_{ub}|$ contributes only at the
level of 10\up{-5}. The inclusive semileptonic decay rates for \keiii\ and
\kmuiii\ are given by
\begin{equation}
\Gamma(K_{\ell3}(\gamma))=|V_{us}|^2\:\left({G_F^2\:M_K^5\over768\pi^3}\right)\:|f_+(0)|^2\:
I_\ell\:S_{\rm EW}\:(1+\delta_\ell),
\label{eq:Gammai}
\end{equation}
where $\ell$ stands for $e$ or $\mu$, and the numerical factor in parentheses
is chosen
so that the phase space integral $I_\ell$ is unity when all final state 
particles 
are massless and there is no $t$ dependence of the form factor. 
The partial widths
$\Gamma(K_{\ell3})$ are given by BR$(K_{\ell3})/\tau_{\kl}$.
For $\tau_{\kl}$, we use the average of the value from 
Eq.~(\ref{eqn:lifetime}),
$\tau_{\kl} = 50.72 \pm 0.36$~ns, and our direct measurement 
\cite{KLOE:KLlife}, $\tau_{\kl} = 50.92\pm0.30$~ns. These two measurements 
are uncorrelated; their average is 
\begin{equation}
\tau_{\kl}=50.84\pm0.23\ {\rm ns}.
\label{eq:tauklave}
\end{equation}
Furthermore, in Eq.~(\ref{eq:Gammai}), $S_{EW}$ \cite{asir} is the 
short-range radiative correction factor, 1.022, 
$f_+(0)=f_+^{K_0}(t=0)$ is the normalization of the vector (and
scalar) form factor at $t=0$ for $K_0\to\pi^+$, and $\delta_\ell$ is the
remaining, mode-dependent, long-range electromagnetic 
correction, \cite{KLOE:gatti}.
The phase space integral $I_{\ell}$ is a function of the form factor
parameters $\lambda_+'$, $\lambda_+''$, and $\lambda_0$. 


We use the values $\lambda_+'=0.0221\pm0.0011$,
$\lambda_+''=0.00023\pm0.0004$, and $\lambda_0=0.0154\pm0.0008$, obtained
from a combined fit to \keiii\ and \kmuiii\
results from KTeV \cite{ktev} and ISTRA+\cite{istra_mu,istra},
weighting the errors as $\sqrt N$.
We obtain:
\begin{eqnarray*}
f_{+}^{K^0}\x\Vus=&0.21638\pm0.00067~\mbox{from}~\keiii,\\
f_{+}^{K^0}\x\Vus=&0.21732\pm0.00087~\mbox{from}~\kmuiii,
\end{eqnarray*}
with an average
\begin{displaymath}
f_{+}^{K^0}\x\Vus=0.21673\pm0.00059
\end{displaymath}
taking correlations into account that mostly arise from the fact that the same
lifetime value enters into the expression for the partial width 
for each channel.
A precise estimate of $f_{+}^{K^0}(0)$, $0.961\pm0.008$,
was first given in 1984 \cite{f0theo}.
Very recently, lattice calculations \cite{latt} have given
the value $f_{+}^{K^0}(0)=0.960\pm0.009$, in excellent agreement
with that from Ref.~\citen{f0theo}. Using the value from Ref.~\citen{f0theo}, 
we find:
\begin{displaymath}
\Vus=0.2257\pm0.0022.
\end{displaymath}
This result can be compared with the value 0.2265$\pm$0.0021
obtained from unitarity if $V_{ud}=0.9740\pm0.0005$~\cite{Czar}.


\section*{Conclusions} 
We have obtained new, precise values of the dominant \kl\ branching
ratios and of the
\kl\ lifetime. Our values for the branching ratios are fully inclusive 
of final-state radiation. Results for the value of the product $V_{us}\x
f_+^{K^0}(0)$ show that the violation of unitarity in the first row of
the CKM matrix suggested by the value for $\mbox{BR}(\keiii)$ in the PDG
listings \cite{PDG} does not exist.
We eagerly await improved estimates of the value of the form factor at
$t=0$.
Better knowledge of the form factor parameters would also be welcome.

\begin{ack}
We gratefully acknowledge the help of F.~Mescia and G.~Isidori
with various aspects of the
determination of $V_{us}\x f_+^{K^0}(0)$ presented in this paper. 
In particular, we thank G.~Isidori for useful suggestions concerning the
simulation of final-state radiation. 
We thank the \DAF\ team for their efforts in maintaining low-background 
running conditions and their collaboration during all data-taking. 
We would like to thank our technical staff: 
G.F.~Fortugno for his dedicated work to ensure efficient operations of 
the KLOE Computing Center; 
M.~Anelli for his continuous support to the gas system and the safety of
the detector; 
A.~Balla, M.~Gatta, G.~Corradi, and G.~Papalino for the maintenance of the
electronics;
M.~Santoni, G.~Paoluzzi, and R.~Rosellini for the general support to the
detector; 
C.~Piscitelli for his help during major maintenance periods.
This work was supported in part by DOE grant DE-FG-02-97ER41027; 
by EURODAPHNE, contract FMRX-CT98-0169; 
by the German Federal Ministry of Education and Research (BMBF) contract 06-KA-957; 
by Graduiertenkolleg `H.E. Phys. and Part. Astrophys.' of Deutsche Forschungsgemeinschaft,
Contract No. GK 742; 
by INTAS, contracts 96-624, 99-37; 
and by the EU Integrated Infrastructure
Initiative HadronPhysics Project under contract number
RII3-CT-2004-506078.
\end{ack}

\bibliographystyle{elsart-num}
\bibliography{kl-vus-6}

\end{document}